\documentclass[twocolumn,twocolappendix]{aastex631}

\usepackage{multirow}
\usepackage{booktabs}

\usepackage{array, threeparttable}

\begin{document}

\title{Dwarf galaxies with the highest concentration are not thicker than ordinary dwarf galaxies}

\author{Lijun Chen }
\email{hzhang18@ustc.edu.cn}
\affiliation{CAS Key Laboratory for Research in Galaxies and Cosmology, Department of Astronomy, University of Science and Technology of China, Hefei 230026, China}
\affiliation{School of Astronomy and Space Sciences, University of Science and Technology of China, Hefei, 230026, China}

\author[0000-0003-1632-2541]{Hong-Xin Zhang}
\altaffiliation{Corresponding author}
\affiliation{CAS Key Laboratory for Research in Galaxies and Cosmology, Department of Astronomy, University of Science and Technology of China, Hefei 230026, China}
\affiliation{School of Astronomy and Space Sciences, University of Science and Technology of China, Hefei, 230026, China}

\author[0000-0001-8078-3428]{Zesen Lin}
\affiliation{Department of Physics, The Chinese University of Hong Kong, Shatin, N.T., Hong Kong S.A.R., China}
\affiliation{ Deep Space Exploration Laboratory/Department of Astronomy, University of Science and Technology of China, Hefei, 230026, People’s Republic of China}
\affiliation{School of Astronomy and Space Sciences, University of Science and Technology of China, Hefei, 230026, China}

\author[0000-0002-4742-8800]{Guangwen Chen}
\affiliation{CAS Key Laboratory for Research in Galaxies and Cosmology, Department of Astronomy, University of Science and Technology of China, Hefei 230026, China}
\affiliation{School of Astronomy and Space Sciences, University of Science and Technology of China, Hefei, 230026, China}

\author{Bojun Tao}
\affiliation{CAS Key Laboratory for Research in Galaxies and Cosmology, Department of Astronomy, University of Science and Technology of China, Hefei 230026, China}
\affiliation{School of Astronomy and Space Sciences, University of Science and Technology of China, Hefei, 230026, China}

\author[0000-0002-2384-3436]{Zhixiong Liang}
\affiliation{School of Mathematics and Physics, Anqing Normal University, Anqing 246011, People's Republic of China}

\author[0000-0003-4959-1625]{Zheyu Lin}
\affiliation{ Deep Space Exploration Laboratory/Department of Astronomy, University of Science and Technology of China, Hefei, 230026, People’s Republic of China}
\affiliation{School of Astronomy and Space Sciences, University of Science and Technology of China, Hefei, 230026, China}

\author[0000-0002-7660-2273]{Xu Kong}
\affiliation{ Deep Space Exploration Laboratory/Department of Astronomy, University of Science and Technology of China, Hefei, 230026, People’s Republic of China}
\affiliation{School of Astronomy and Space Sciences, University of Science and Technology of China, Hefei, 230026, China}
\affiliation{Frontiers Science Center for Planetary Exploration and Emerging Technologies, University of Science and Technology of China, Hefei, Anhui, 230026, China}

\begin{abstract}

{The formation mechanism of high-concentration dwarf galaxies is still a mystery.} We perform a comparative study of the intrinsic shape of nearby low-mass galaxies with different stellar concentration.\ The intrinsic shape is parameterized by the intermediate-to-major axis ratios B/A and the minor-to-major axis ratios C/A of triaxial ellipsoidal models. Our galaxies ($10^{7.5} M_\odot$ $\textless$ $M_\star$ \textless $10^{10.0} M_\odot$) are selected to have spectroscopic redshift from SDSS or GAMA, and have  broadband optical images from the HSC-SSP Wide layer survey. The deep HSC-SSP images allow to measure the apparent axis ratios $q$ at galactic radii beyond the central star-forming area of our galaxies. We infer the intrinsic axis ratios based on the $q$ distributions. We find that 1) our galaxies have typical intrinsic shape similarly close to be oblate ($\mu_{B/A}$ $\sim$ 0.9--1), regardless of the concentration, stellar mass, star formation activity, and local environment (being central or satellite); 2) galaxies with the highest concentration tend to have intrinsic thickness similar to or (in virtually all cases) slightly thinner (i.e. smaller mean $\mu_{C/A}$ or equivalently lower triaxiality) than ordinary galaxies, regardless of other properties explored here. This appears to be in contrast with the expectation of the classic merger scenario for high-concentration galaxies. Given the lack of a complete understanding of dwarf-dwarf merger, we cannot draw a definite conclusion about the relevance of mergers in the formation of high-concentration dwarfs. Other mechanisms such as halo spin may also play important roles in the formation of high-concentration dwarf galaxies. 

\end{abstract}

\keywords{dwarf galaxies --
            intrinsic shape --
            high-concentration --
            color --
            environment}

\section{Introduction} \label{sec:intro}
   \par
   Dwarf galaxies (stellar mass $\lesssim 5\times 10^{9} M_{\odot}$) are the most common galaxies in the universe, and include a variety of morphological types such as dwarf irregular galaxies, dwarf ellipticals, blue compact dwarfs, compact ellipticals, ultra-fiant dwarfs, ultracompact dwarfs, etc. The morphology of galaxies are known to be highly correlated with properties such as mass, color and environment (e.g.,  \citealt{2007MNRAS.382.1187K, 2020ApJ...900..163K}). Understanding the formation mechanisms of dwarfs with different morphological types and their possible evolutionary connection is an important part of a complete theory of galaxy formation and evolution.

   \begin{figure*}[htp]
   \centering
   \includegraphics[width=15cm]{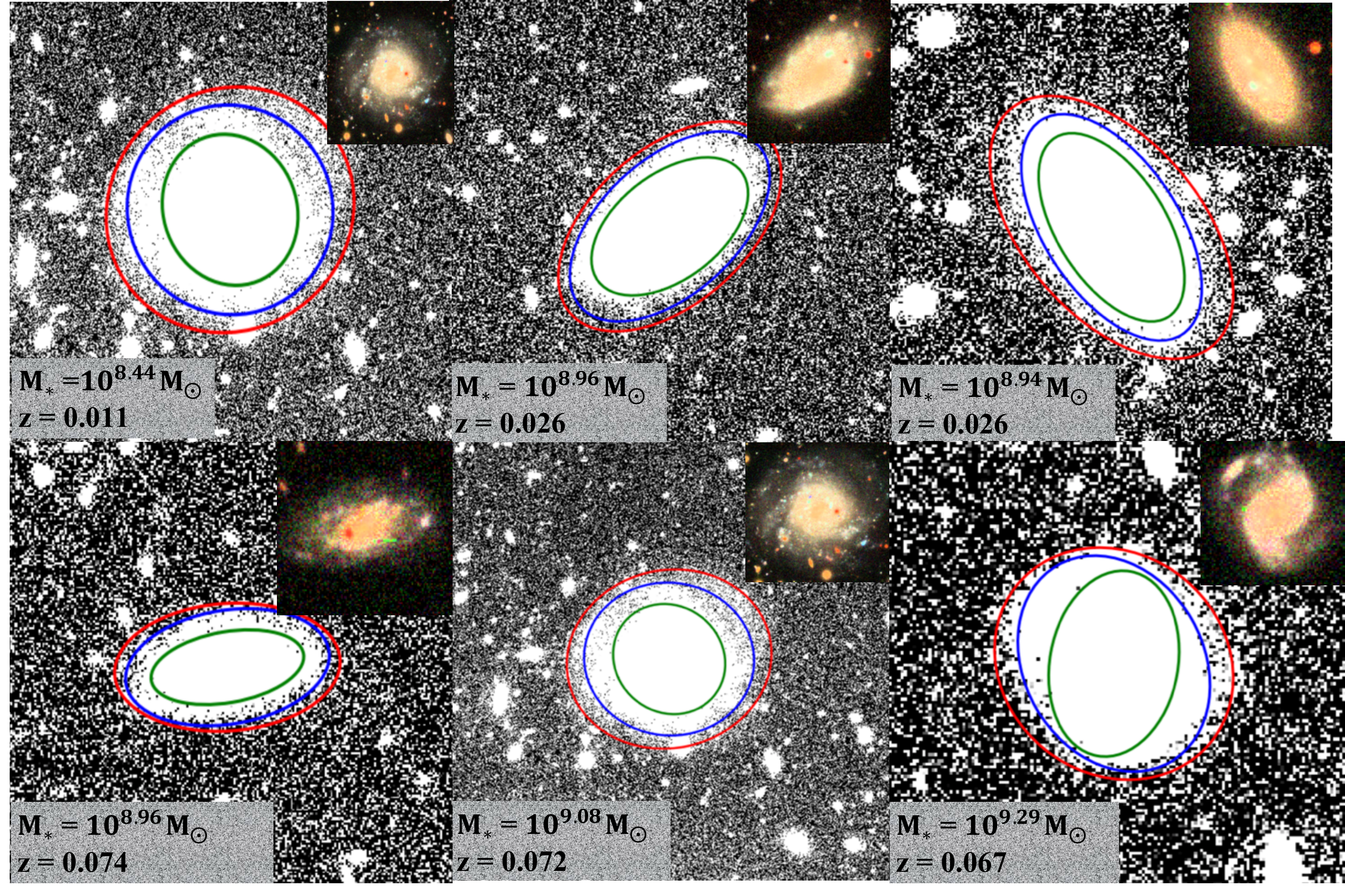}
   \caption{HSC optical images of six dwarf galaxies randomly picked from our sample. The top(bottom) row show galaxies at the low- (high-) redshift end of the sample. In each panel, the main figure is the i-band grayscale image, and the inset figure is the $i$-, $r$- and $g$- band composite false color RGB image. The i-band isophotal ellipses at 25 (green), 26 (blue), and 27 (red) mag/$\rm arcsec^{2}$ are overplotted on each main figure. The color images are created with the \cite{2004PASP..116..133L} method as implemented in ${\rm astropy.visualization.make_lupton_rgb}$. The redshift and stellar mass of each galaxy are indicated at the lower left corner of each panel.}
   \label{fig:rbg}
   \end{figure*}     
   \par
  Generally speaking, the stellar and gaseous structures of galaxies may be shaped by a variety of intrinsic factors(e.g., star formation, mass, spin and concentration of the host dark matter halos) and/or environmental factors (e.g., gravitational and hydrodynamic interactions, gas accretion from cosmic web). A long-standing debate is the origin of compact or high-concentration galaxies and whether they have an evolutionary connection with the far more numerous ordinary galaxies. Some of the commonly cited mechanisms, such as violent disk instability (e.g., \citealt{2009ApJ...703..785D, 2008ApJ...688...67E}), misaligned gas infall (e.g., \citealt{2015MNRAS.450.2327Z, 2009MNRAS.396..696S}), accretion (e.g., \citealt{2018MNRAS.480.2266M, 2018MNRAS.478.3994C}), and galaxy mergers (e.g., \citealt{1977egsp.conf..401T, 1983MNRAS.205.1009N}) are thought to be able to change the gaseous and stellar structures. Particularly, it has been well-established that mergers of spiral galaxies tend to leave remnants with much higher concentration than their progenitor disk galaxies (e.g., \citealt{2012ApJ...756...26M, 2014MNRAS.441.3679A}). The well-known environment-morphology correlation of galaxies \citep{1980ApJ...236..351D}, together with other observational evidences of more vigorous gravitational and hydrodynamic interactions in higher density environment, suggest that galaxy morphology may be affected or re-shaped by environment. 

\begin{figure*}[htp]
    \centering
    \includegraphics[width=18cm]{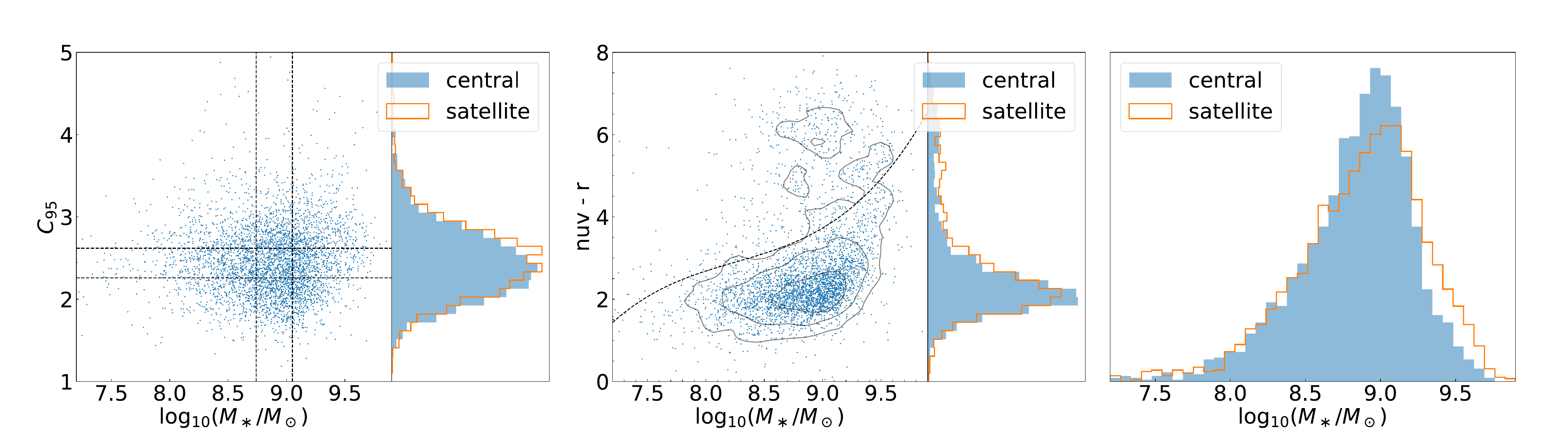}
    \caption{Distribution of the relevant global properties of our sample galaxies. \emph{Left}: Mass-concentration index ($C_{95}$) scatter plot of the sample (main figure), together with the marginal histograms of $C_{95}$ for the central and satellite subsamples respectively. The dashed black lines in the main figure indicate the mass and $C_{95}$ intervals used for subsample definition. \emph{Middle}: mass--(NUV-r) color scatter plot of the whole sample (main figure), together with marginal histograms of (NUV-r) for the central and satellite subsamples. The dashed black line indicate the demarcation line of red quiescent dwarf and blue star-forming dwarf galaxies. The iso-density contours in the main figure enclose the central $34\%$, $68\%$, and $95\%$ of the whole sample. Note that the demarcation between red and blue subsamples is chosen to be approximately follow the upper limit of the $95\%$ contour in the main figure. \emph{Right}: histograms of the stellar mass of central and satellite galaxies.}
    \label{fig:scatter}
\end{figure*}


   \par
    As with their high-mass counterpart, one popular scenario put forward for the formation of high-concentration dwarfs is dwarf-dwarf mergers (e.g., \citealt{2008MNRAS.388L..10B, 2016MNRAS.462.3314W, 2016A&A...595A..56S}). However, unambiguous observational evidence for this merger scenario is still lacking for dwarf galaxies. Particularly, a recent study by \cite{Zhang2020b} casts doubt on the relevance of mergers for the formation of compact or high-concentration dwarfs.  A general lesson learned from numerical simulations is that the remnants of galaxy mergers (especially those between galaxies of comparable masses) tend to become more triaxial or thicker in intrinsic shapes than their progenitors (e.g., \citealt{2006ApJ...650..791C, 2021A&A...647A..95P}). It is thus possible to test the importance of merger scenario through a statistical analysis of the intrinsic shapes of dwarf galaxies with different concentration.

   \par
   The average intrinsic shape of a population of galaxies can be inferred from the distribution of their apparent axis ratios, by assuming that the galaxies can be approximated by a family of optically-thin triaxial ellipsoids (see Section \ref{sec:infershape} for details). This method has been extensively used in the literature to study intrinsic shape of galaxies of different morphological types and masses (e.g., \citealt{1989AJ.....97.1600I,1989ApJ...346L..53F,1998ApJ...499..140S,1998ApJ...505..199S,2010MNRAS.406L..65S,2017ApJ...838...93B,2019MNRAS.486L...1S, 2018ApJ...863L..19L,2020ApJ...899...78R,2020ApJ...900..163K}). To mention several studies relevant to dwarf galaxies, \cite{1998ApJ...499..140S} and \cite{1998ApJ...505..199S} studied the intrinsic shapes of a sample of $\sim$ 170 nearby dwarf galaxies, and found that dwarf irregulars and blue compact dwarfs (BCDs) seem to have similar intrinsic shapes, and dwarf elliptical galaxies are best described as thick oblate spheroid with a mean axial ratio of 0.6; \cite{2020ApJ...900..163K, 2021ApJ...920...72K} found that blue dwarfs and red dwarfs have little difference in their outskirts. \cite{2019MNRAS.486L...1S} found that dwarfs at different environment show little difference in the intrinsic shapes.
    
   \par 
   In this work, we perform a comparative analysis of low-mass galaxies with different stellar concentration, masses, colors and environments, in order to shed light on the formation mechanisms of galaxies with different stellar structures. We select a large sample of local dwarf galaxies from the Sloan Digital Sky Survey (SDSS, \citealt{2000AJ....120.1579Y}) spectroscopic survey and the Galaxy and Mass Assembly (GAMA, \citealt{2011MNRAS.413..971D}) spectroscopic survey. We measure the apparent axis ratios of these galaxies based on exquisite optical images from the Hyper Suprime-Cam Subaru Strategic Program (HSC-SSP, \citealt{2018PASJ...70S...4A}) {data release 2} {(DR2)}. The rest of the paper is organized as follows. Section \ref{sec:sample} describes the sample selection and data. Section \ref{sec:analysis} describes the measurement of apparent axis ratios and other parameters used in this work, and the inference of the intrinsic axis ratios. Sections \ref{sec:result} and \ref{sec:discussion} present our result and discussion, respectively. Section \ref{sec:summary} summarizes our main conclusion. Throughout this paper, we assume a standard flat {$\Lambda$}CDM cosmology with $H_0$ = 70 $\rm km$ $\rm s^{-1}$ $\rm Mpc^{-1}$ and $\Omega_m$ = 0.3.

\begin{figure*}[htp]
    \centering
    \includegraphics[width=15cm]{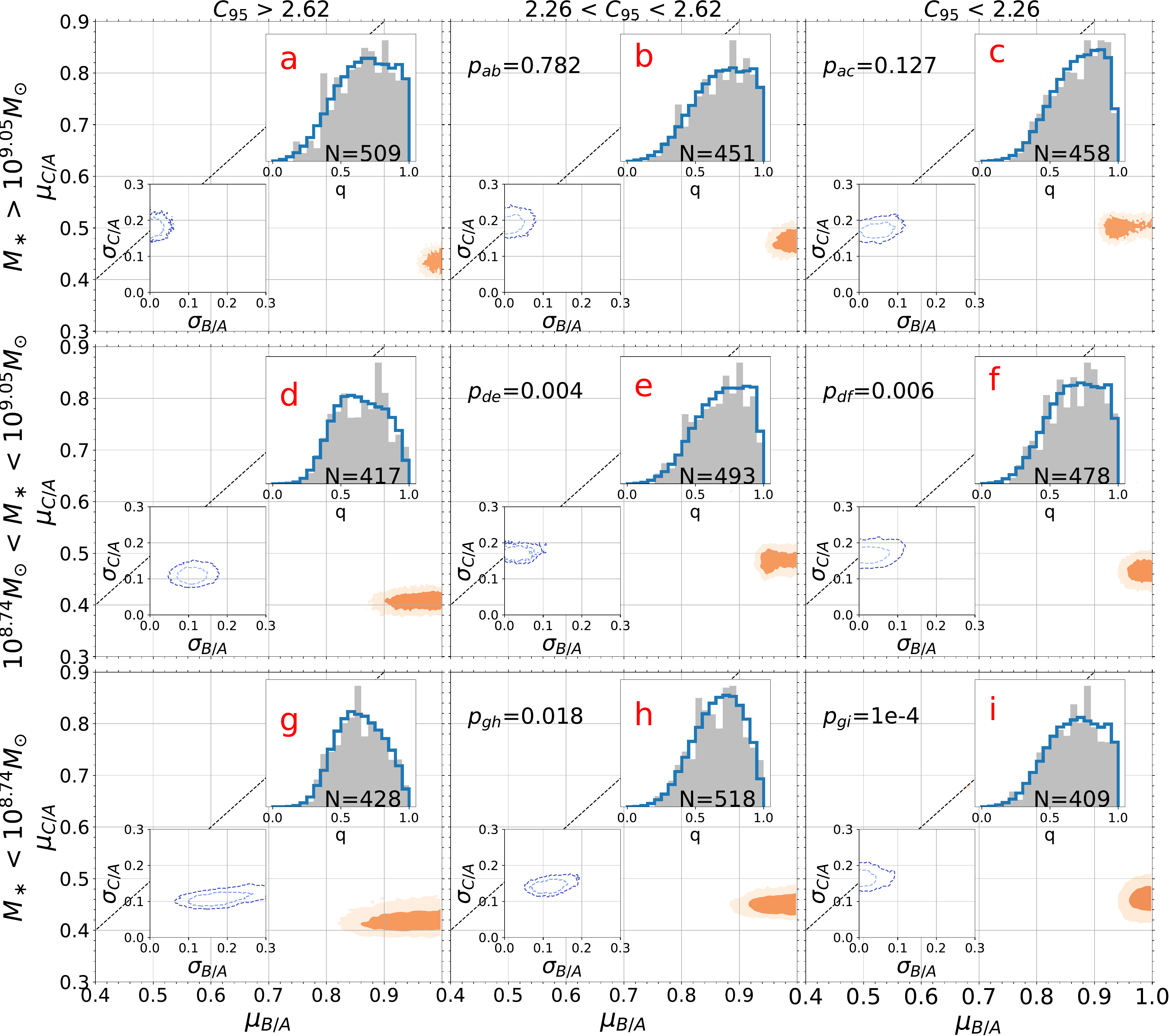}
    \caption{The intrinsic mean axis ratios $\mu_{B/A}$ and $\mu_{C/A}$ inferred for subsamples with different stellar mass (rows) and concentration indices $C_{95}$ (columns). In the main plot of each panel, the red and pink filled contours respectively represent the central 68\% and 95\% inter-percentile range of the Bayesian posterior distributions of $\mu_{B/A}$ and $\mu_{C/A}$.\ The black dashed line indicates the B = C line. {The inset plot at the bottom of each panel shows the central 68\%  and 95\% inter-percentile range of the posterior distributions of $\sigma_{B/A}$ (x axis) and $\sigma_{C/A}$ (y axis).} The inset plot at the top of each panel shows the observed axis ratio $q$ distribution (filled grey histogram) and a model distribution obtained by randomly sampling viewing angles of ellipsoids characterized by the most likely mean intrinsic axis ratios and standard deviations (open blue histogram). The number of galaxies in each subsample is indicated at the upper right corner of each panel. {The p-values of K-S test for the $q$ distributions of the highest-$C_{95}$ (panel ID: a, d, g) and lower-$C_{95}$ subsamples in each mass interval are also indicated at the top right corner of each panel.}}
    \label{fig:MASS_CI}
\end{figure*}

\section{Data and Sample Selection} \label{sec:sample}
   
    \par
    Measurements of apparent axis ratios may be affected by spatial irregularities of the distribution of bright star-forming regions in galaxies. In order to measure axis ratios of the underlying stellar mass distribution of galaxies, deep images that reach the relatively quiescent outskirts of galaxies are necessary. To this end, we use the high-quality optical images ($g, i, r$) from HSC-SSP DR2. The wide-layer survey of HSC-SSP covers a sky area of $\sim$ 1,100 square degrees, with a median $i$-band seeing of $\sim$ 0.6 arcsec, and reaches a surface brightness sensitivity $\textgreater$ 28 mag/$\rm arcsec^{2}$ that is 2 $\sim$ 3 mag deeper than the recently completed DESI Legacy Imaging Surveys \citep{2019AJ....157..168D}.
    
    \par
    The parent sample of low-mass galaxies is selected from the Legacy Survey spectroscopic catalogs of SDSS DR16 \citep{2020ApJS..249....3A} and GAMA DR4 \citep{2022MNRAS.513..439D}. The SDSS Legacy Survey covers a total sky area of $\sim$ 8240 square degrees and is nearly compete for galaxies down to 17.77 mag in $r$-band. The GAMA survey covers $\sim$ 286 square degrees and reaches a r-band completeness limit of $\sim$ 19.7 mag. In particular, the parent sample is selected according to the following criteria: 1) 0.005 $\textless$ z $\textless$ 0.2; 2) r-band absolute magnitude {$-$18.5 \textless~$M_{r}$ \textless~$-$14.8 mag}; 3) average surface brightness $\mu_r$ within the half-light radius \textless~24.5 mag/$\rm arcsec^{2}$ for SDSS and \textless~25 mag/$\rm arcsec^{2}$ for GAMA. {We note that the fainter $M_{r}$ limit of $-$14.8 mag is chosen such that the luminosity-volume test \citep{Schmidt1968} gives average $V/V_{\rm max}$\footnote{$V$ is the comoving volume between the galaxy and the observer covered by a given survey, and $V_{\rm max}$ is the comoving volume that a galaxy could be observed given the flux limit of a survey.} $\simeq$ 0.5 across the full luminosity range under consideration, which guarantees a valid $1/V_{\rm max}$ weighting that statistically convert our flux-limited sample to a volume-limited sample to be used in the following analysis}. The criterion of surface brightness is in line with the respective $90\%$ completeness limit of SDSS \citep{2022ApJS..259...35A} and GAMA \citep{2018MNRAS.474.3875B}.
    
    \par
    After matching the parent sample selected from SDSS and GAMA with the HSC-SSP wide-layer imaging survey footprint, we end up with 8136 galaxies. A visual inspection of the images of these galaxies and the isophotal fitting results (see next section) reveals that the outskirts of 3793 galaxies are significantly contaminated by either bright stars or galaxies projected near the line of sight. These contaminated galaxies are excluded from our following analysis.
    
    \par
    Lastly, we retrieve the near-UV (NUV) photometry of our galaxies from the public catalog release of the Galaxy Evolution Explorer (GALEX) satellite. The NUV-optical colors are used to divide our sample into star-forming and quiescent subsamples (Figure \ref{fig:scatter}).

\begin{figure*}[htp]
    \centering 
    \includegraphics[width=10cm]{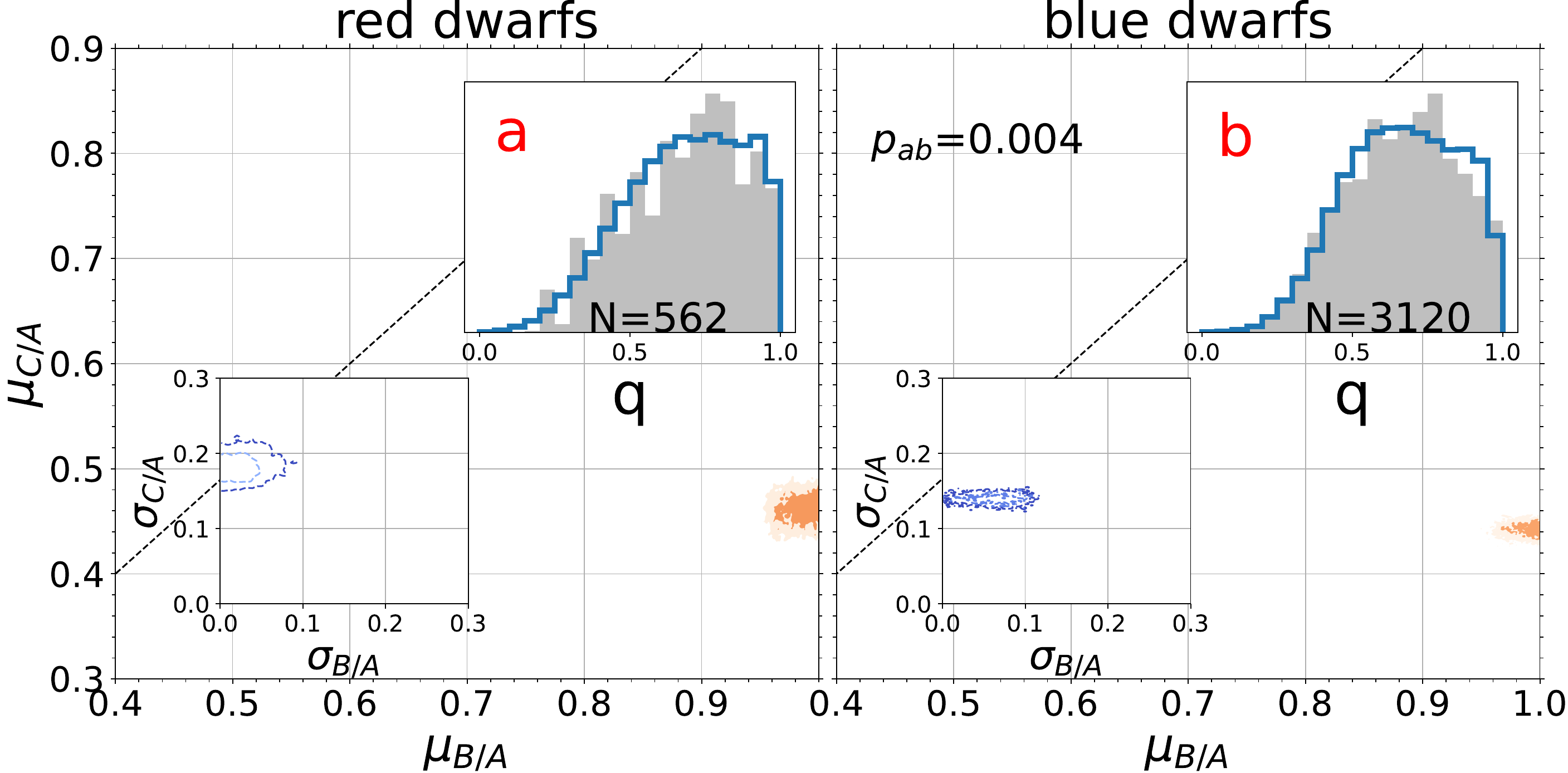}
    \caption{Same as Figure \ref{fig:MASS_CI} but for red dwarfs(left) and blue dwarfs(right).}
    \label{fig:color}
\end{figure*}

\begin{figure*}[htp]
    \centering
    \includegraphics[width=15cm]{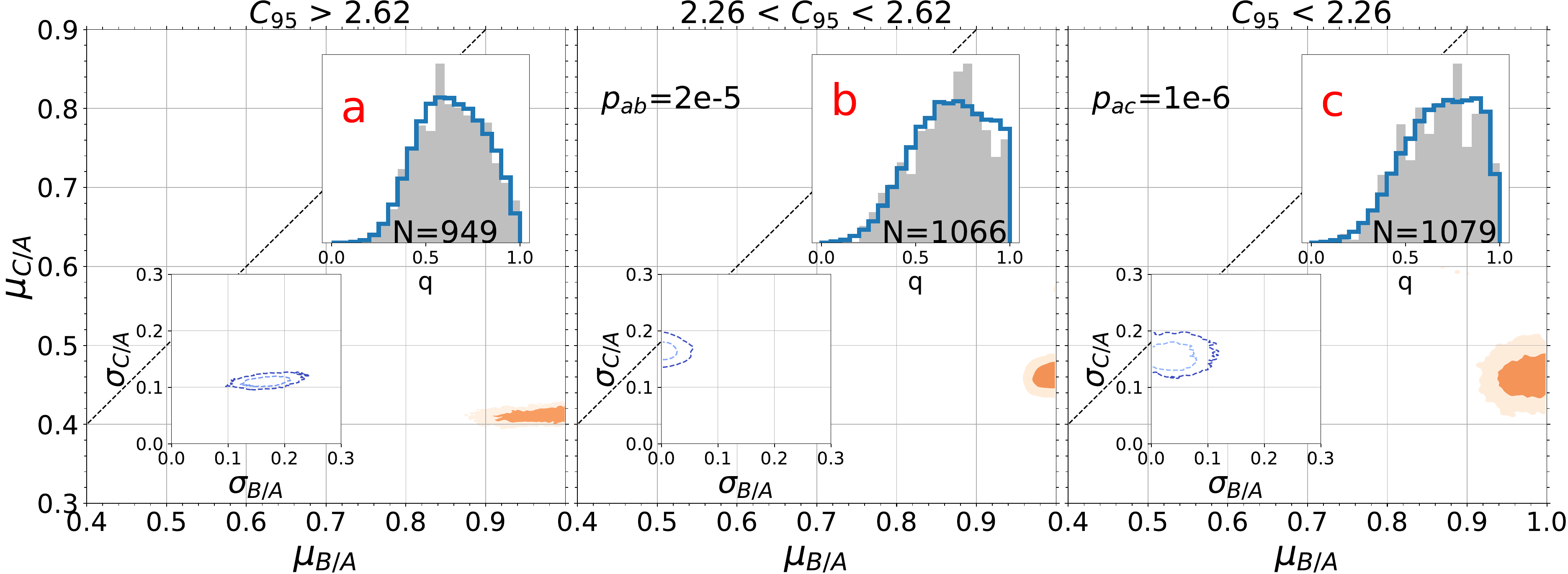}
    \caption{Same as Figure \ref{fig:MASS_CI} but for for blue dwarfs with different concentraion indices $C_{95}$ . }
    \label{fig:blue_CI}
\end{figure*}  


\section{Analysis}\label{sec:analysis}

   \subsection{Measurement of Apparent Axis Ratios}
   \par
   The axis ratios $q$ (=b/a) are determined by fitting {$i$-band} galaxy light distributions with the \textsc{IRAF} task ELLIPSE \citep{1987MNRAS.226..747J}. To avoid influence from nearby objects, we use masks generated by Sextractor \citep{1996A&AS..117..393B} when running Ellipse fitting. After running ELLIPSE in non-interactive mode for the whole sample, we perform a visual inspection by overplotting the best-fit ellipses on isophotal contours of galaxy images. For fittings that failed to converge, we re-run ELLIPSE by adjusting the initial parameters (e.g., center, starting semi-major axis length, ellipticity, and position angle) until the fitting converges. All of the ELLIPSE geometry parameters are allowed to vary with radius during the fitting. 
   
   \par 
   Reliable ELLIPSE geometry parameters are measured out to galactocentric distances with surface brightness of $\gtrsim$ 27 mag/$\rm arcsec^{2}$. {To reduce random errors in the measurement, we adopt the mean values determined at the last three radial annuli where {the $i$-band surface brightness is $\leq$ 27 mag/arcsec$^2$}}.\ Images of some example galaxies overplotted with i-band isophotal ellipses are showed in Figure \ref{fig:rbg}.\ {It is worth mentioning that our motivation of choosing geometry parameters determined at galactocentric distances of similarly faint surface brightness levels, instead of a fixed factor of (say) half-light radius, is twofold: 1) to avoid the influence of irregular spatial distribution of star formation close to the inner region of low-mass galaxies, and 2) to avoid potential systematic bias for galaxies with different concentration. With that in mind, we note that our conclusion in this work does not change if choosing the geometry parameters measured at sufficiently large factors (e.g. $>$ 3$\times$) of half-light radius.}

\begin{figure}[htp]
    \centering
    \includegraphics[width=8cm]{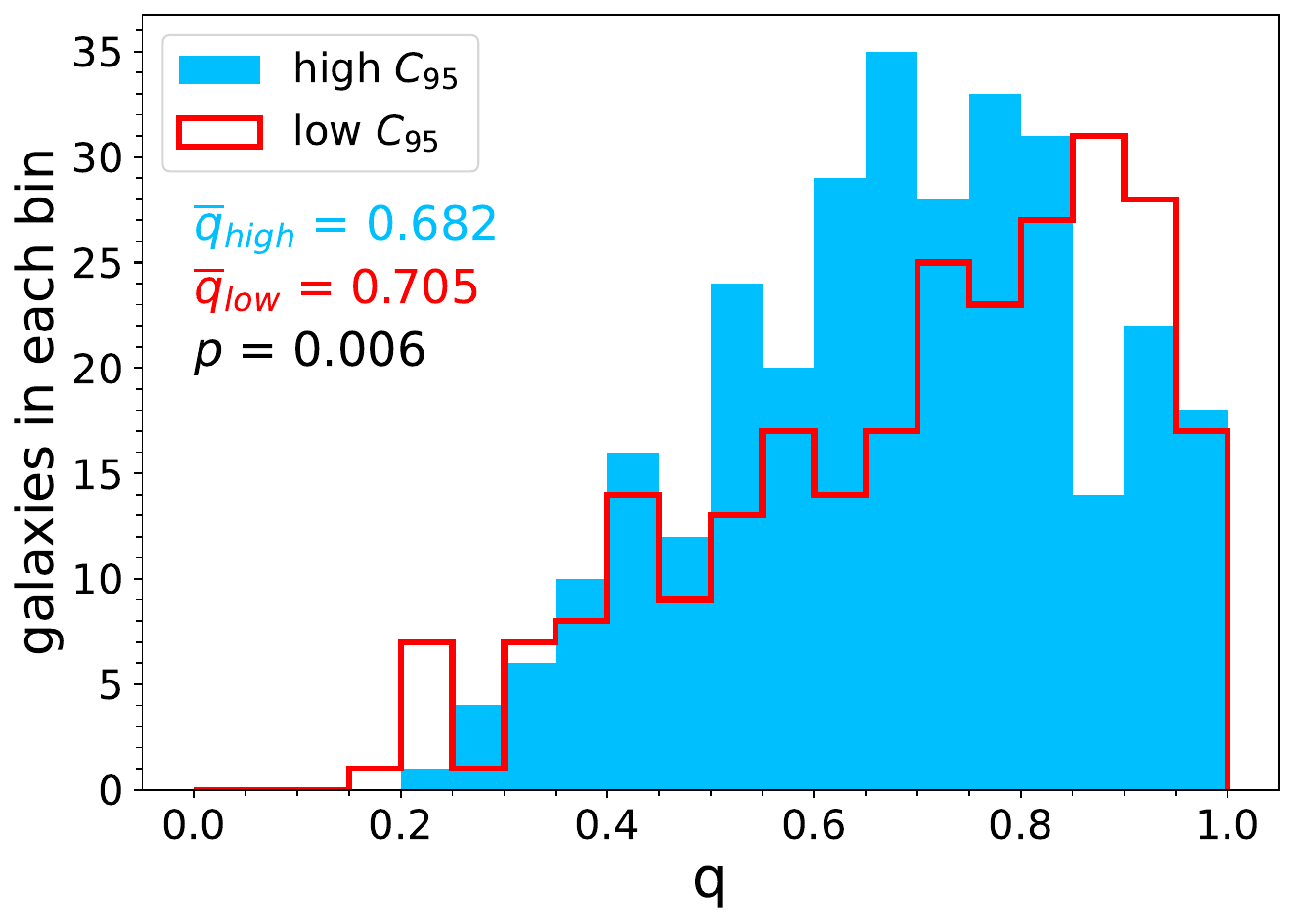}
    \caption{Histogram of the observed axis ratios of red dwarfs with different $C_{95}$.\ Filled histogram is for high-concentration red dwarf subsample, and open histogram is for low-concentration red dwarf subsample. Note that the distribution of low-$C_{95}$ red dwarfs is skewed toward larger average axis ratios.\ The K-S test p-value for the two subsamples is indicated in the figure.}
    \label{fig:red}
\end{figure} 

\begin{figure*}[htp]
    \centering
    \includegraphics[width=15cm]{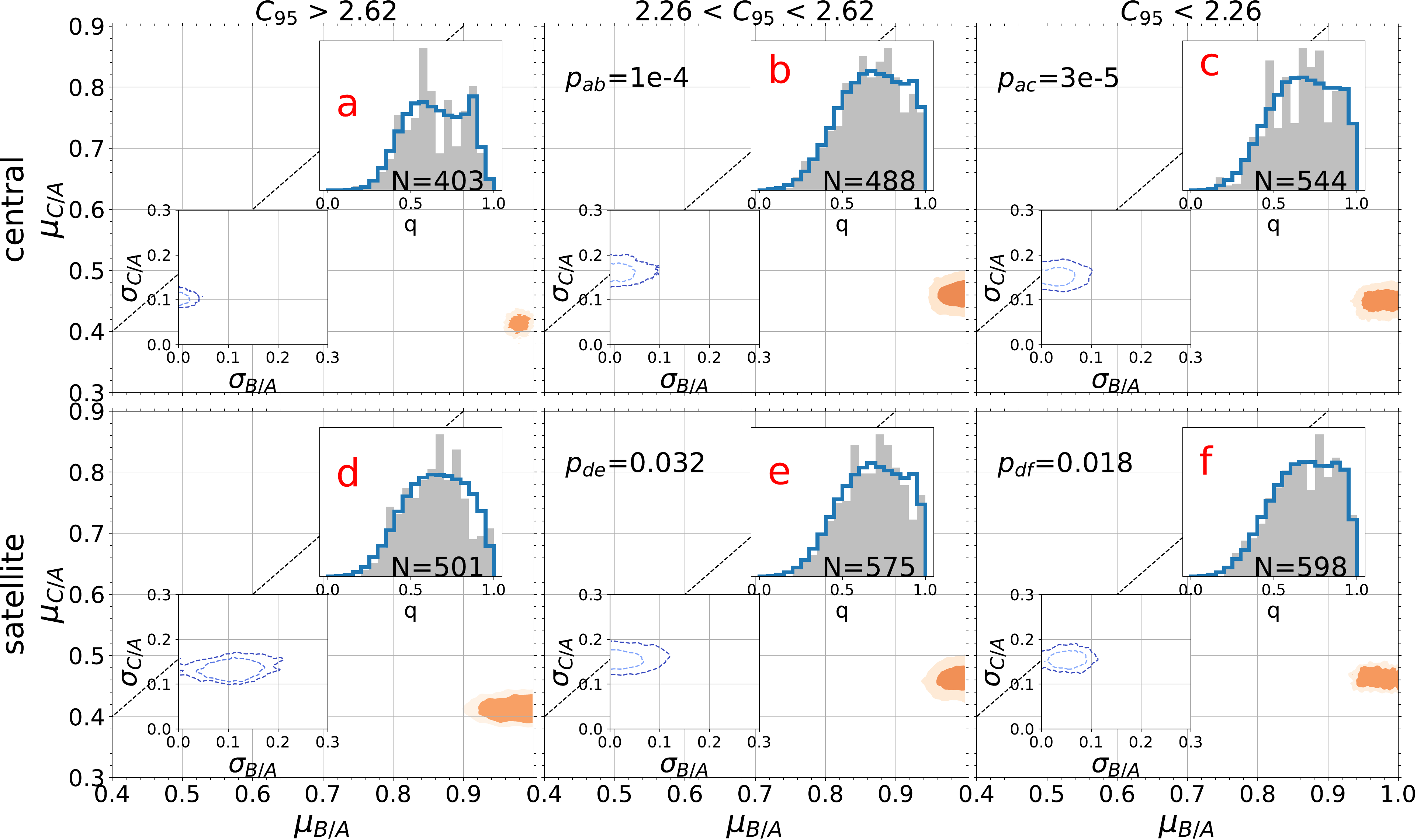}
    \caption{Same as Figure \ref{fig:MASS_CI} but for blue dwarfs in different environments (rows) with different concentraion indices $C_{95}$ (columns).}
    \label{fig:blue_envirCI}
\end{figure*}

  \subsection{Inference of Intrinsic Shape}\label{sec:infershape}
  \par
  Following previous studies\cite[especially][hereafter K20]{2020ApJ...900..163K}, we assume that our galaxies can be described by a family of optically-thin triaxial ellipsoids. The ellipsoids are characterized by the intrinsic ratios of their major (A), intermediate (B), and minor (C) axes. Under this assumption, the observed axis ratios $q$ (b/a=1-ellipticity) represent a projection of the ellipsoids on the sky. There is no one-to-one correspondence between $q$ and the intrinsic axis ratios (i.e., B/A and C/A). However, by projecting large samples of ellipsoids at random viewing directions, as quantified by the polar angle $\theta$ and azimuthal angle $\phi$, ellipsoids of different intrinsic axis ratios give distinct distributions of apparent axis ratios.
  
  \par
  We infer the intrinsic mean axis ratios for subsamples of our galaxies through the Bayes's theorem. In particular, following K20, assuming B/A and C/A follow normal distributions, a Poisson likelihood function is adopted to compare the observed distribution of axial ratios $q$ with that predicted (with a uniform random sampling of $\theta$ and $\phi$) by a given set of triaxial ellipsoid parameters: $\mu_{B/A}$, $\sigma_{B/A}$, $\mu_{C/A}$ and $\sigma_{C/A}$, where $\mu_{B/A}$ is the mean of B/A, $\sigma_{B/A}$ is the standard deviation of B/A, $\mu_{C/A}$ is the mean of C/A, and $\sigma_{C/A}$ is the standard deviation of C/A. A flat prior distribution from 0 to 1 is assumed for $\mu_{B/A}$ and $\mu_{C/A}$, and from 0 to 0.5 for $\sigma_{B/A}$ and $\sigma_{C/A}$.  We use the Markov Chain Monte Carlo (MCMC) method implemented in the $\tt emcee$ package \citep{2013PASP..125..306F} to sample the posterior distribution of the four ellipsoid parameters. We note that a bin size of 0.05 is used to construct $q$ distributions for both observations and model predictions. Thanks to the relatively large sample size, our results are not sensitive to the exact choice of bin size. 

  \subsection{Determination of Concentration Index and Stellar Mass}
  Concentration indices (C=$R_{90}/R_{50}$; hereafter $C_{95}$) of galaxies are determined based on elliptical aperture photometry {in the $i$ band}, where $R_{90}$ and $R_{50}$ are respectively the semi-major axis radius containing $90\%$ and $50\%$ of the Petrosian flux. We note that, unlike the concentration indices calculated based on circular aperture photometry, elliptical aperture concentration indices are not subject to significant influence of galaxy inclinations \citep{2005AJ....130.1545Y}. 
  
  Stellar mass of galaxies in the final sample is estimated using the color--mass-to-light-ratio relations calibrated for local group dwarf galaxies \citep{2017ApJS..233...13Z}.\ Our galaxies span a stellar mass range of 7.5 $\textless$ log($M_*$/$M_{\odot}$) $\textless$ 10.
 
  \subsection{Classification of Central and Satellite Galaxies}
  To explore the environmental dependence of galaxy shapes, we classify our galaxies into central and satellite subsamples. Specifically, a galaxy is classified as a central galaxy if it does not fall into the (projected) virial radii and $\pm$ 3 times the virial velocities of any brighter galaxies.\ {To perform the classification, both the SDSS and GAMA spectroscopic catalogs and the SDSS Value-Added Photo-z catalog of \cite{Cunha2009} are used.\ Particularly, if a galaxy does not have  spectroscopic redshift measurement but has a $>$ 10\% probability of sharing the same redshift as the galaxy in question in our sample, it is considered as a real neighbour to the galaxy in our sample} \citep[see also][for a similar selection strategy]{Wang2019}.\ Note that the virial radii and virial velocities of host dark matter halos of our galaxies are estimated from their stellar mass through abundance matching, as calibrated by  \cite{2010MNRAS.404.1111G}.


\section{Results: Intrinsic Shape of Galaxies with Different Concentration}
\label{sec:result}
  \par
  We perform a comparative analysis of the intrinsic shapes of dwarf galaxies with different concentration indices $C_{95}$. $C_{95}$ measures the overall curvature of the radial distribution of light, and is positively correlated with the S\'{e}rsic index (e.g., \citealt{2001AJ....121.2358B}). A higher $C_{95}$ means a stronger light concentration toward the center and a more extended wings toward the outskirts. Our galaxies are classified into three approximately equally sized subsamples based on $C_{95}$:  high-concentration ($C_{95}$ $\textgreater$ 2.62), intermediate-concentration (2.62 $\textgreater$ $C_{95}$ $\textgreater$ 2.26), and low-concentration ($C_{95}$ $\textless$ 2.26) subsamples.\ {Our experiment (not shown in the paper) suggests that the conclusion in this work is not sensitive to the exact choice of the division lines of $C_{95}$.}
  
  \par
  To separate out the effects of galaxy properties other than concentration, the concentration parameter is paired with stellar mass, NUV-$r$ color (as an indicator of light-weighted stellar ages) and environment (central vs. satellite galaxies), respectively, to form different subsamples for our comparative analysis. The relevant results are presented in the following subsections.

  \subsection{Mass Dependence}
  
  \par
  Here the three samples with different $C_{95}$ are further classified into high-mass  (log($M_*$/$M_{\odot}$) $\textgreater$ 9.05), intermediate-mass (9.05 $\textgreater$ log($M_*$/$M_{\odot}$) $\textgreater$ 8.74) and low-mass (8.74 $\textgreater$ log($M_*$/$M_{\odot}$) $\textgreater$ 7.5) galaxies. The demarcation limits of mass are chosen such that the resultant subsamples have about the same number of galaxies. We note that the following results are not sensitive to the exact choice of the demarcation limits.
  
  \par
  Posterior probability distributions of the intrinsic mean axis ratios ($\mu_{B/A}$ and $\mu_{C/A}$) of the above defined six subsamples are presented in Figure \ref{fig:MASS_CI}. The filled contours enclose the central $68\%$ (red color) and $95\%$ (pink color) of the posterior distributions of $\mu_{B/A}$ and $\mu_{C/A}$.\ The central 68\% and 95\% inter-percentile range of $\sigma_{B/A}$ and $\sigma_{C/A}$ are shown as open contours in the inset plot at the bottom of each panel. The results plotted in Figure \ref{fig:MASS_CI} are also listed in Tables \ref{table1} and \ref{table2}.
  
  \par
  All of the subsamples have $\mu_{B/A}$ $\sim$ 0.9 -- 1.0 and $\mu_{C/A}$ $\sim$ 0.4 -- 0.5, suggesting oblate-triaxial intrinsic shapes for these galaxies. {Given the 68\% inter-percentile range}, subsamples of given stellar mass intervals have similar $\mu_{B/A}$, but the high-$C_{95}$ subsamples have systematically smaller $\mu_{C/A}$ than the intermediate- and low-$C_{95}$ subsamples. For the high-$C_{95}$ subsamples, the low- and intermediate-mass subsamples have smaller $\mu_{C/A}$ than the high-mass subsample.\ Regarding the standard deviation of the intrinsic axis ratios, except for the high-mass bin, the high-$C_{95}$ subsamples have  marginally larger $\sigma_{B/A}$ but smaller $\sigma_{C/A}$ than the lower-$C_{95}$ subsamples.
  
  {The intrinsic shape  differences of different subsamples are clearly reflected in the observed $q$ distributions (the inset plot of each panel), in the sense that 1) higher-$C_{95}$ subsamples exhibit a proportional lack of galaxies with round appearance compared to lower-$C_{95}$ subsamples and 2) the low- and intermediate-mass subsamples exhibit a proportional lack of galaxies with round appearance compared to the high-mass subsample. 
  To quantify the significance of the difference between $q$ distributions of subsamples with different $C_{95}$, we perform the K-S test and indicate the $p$-values in Figure \ref{fig:MASS_CI}. Except for the high-mass subsamples, the K-S test suggests a significant difference (p-value $<$ 0.05) between $q$ distributions of the high-$C_{95}$ and intermediate/low-$C_{95}$ subsamples, corroborating the above finding.}

\begin{figure*}[htp]
    \centering
    \includegraphics[width=15cm]{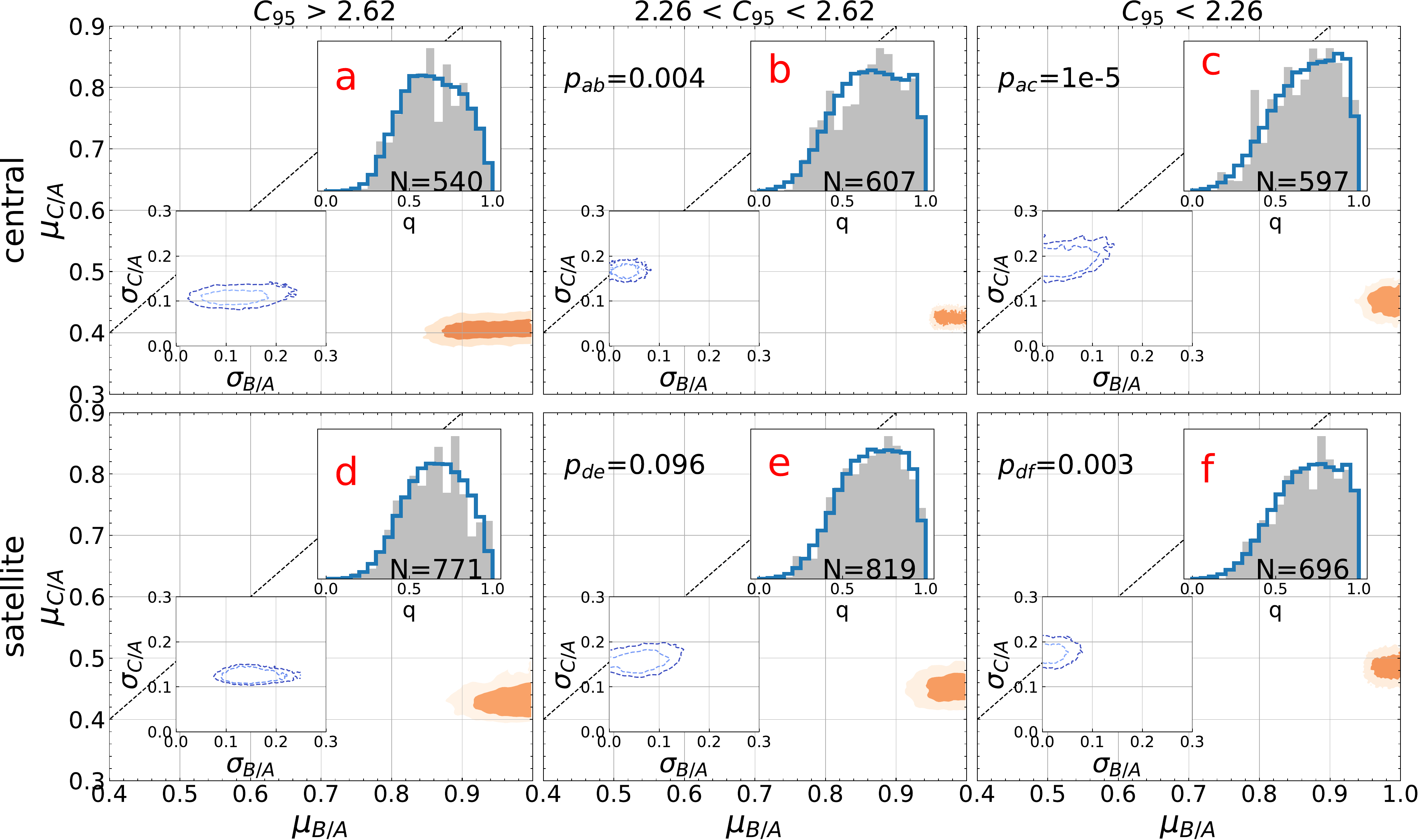}
    \caption{Same as Figure \ref{fig:MASS_CI} but for dwarfs in different local environments (rows) and with different concentration indices $C_{95}$ (columns).}
    \label{fig:envir_CI}
\end{figure*}

  \subsection{Dependence on NUV--$r$ color}
  
  \par
  Here the whole sample is divided into blue and red subsamples, based on the NUV--$r$ vs.\ stellar mass distribution (middle panel of Figure \ref{fig:scatter}).\ The whole  sample is dominated by relatively blue galaxies, peaking near NUV--$r$ $\sim$ 2.0 mag and without an obvious red sequence.\ The somewhat arbitrary demarcation between the blue and red subsamples {(NUV--$r$ = 0.51$M_{\star}^3$ --12.8$M_{\star}^2$ + 106.47$M_{\star}$--259.09)} is chosen to largely follow the upper bound of the contour enclosing the central 95$\%$ of the data points near the blue peak in the NUV--$r$---mass distribution. This demarcation curve effectively separates galaxies that are significantly redder than the numerically dominant ``blue cloud'' galaxies. {Our experiment suggests that the following conclusion would not change if shifting the division line upward or downward by, say, 0.5 mag.}
  
  \par
  Similar to Figure \ref{fig:MASS_CI} for the mass dependence, Figure \ref{fig:color} presents the posterior distributions of $\mu_{C/A}$ and $\mu_{B/A}$ for the two subsamples divided by NUV--$r$.\  Although the two subsamples have similar $\mu_{B/A}$ within the 68\% inter-percentile ranges, the blue subsample has significantly smaller mean $\mu_{C/A}$, and consistently, a $q$ distribution skewed to smaller values than does the red subsample. The K-S test for the $q$ distributions of the two subsamples suggests that the two distributions are significantly different ($p$-value = 0.004). This finding of a flatter shape for blue dwarfs is in line with \citet{2021ApJ...922..267C} where star-forming dwarfs were visually selected based on the presence of star-forming regions in optical images.
  
  \par
  We further explore the $C_{95}$ dependence of the intrinsic shapes of the blue galaxies. The results are presented in Figure \ref{fig:blue_CI} (and also Tables \ref{table1} and \ref{table2}). The high-$C_{95}$ subsample of blue galaxies has significantly smaller mean $\mu_{C/A}$ than the low- and intermediate-$C_{95}$ subsamples. In addition, the high-$C_{95}$ blue subsample has significantly larger mean $\sigma_{B/A}$ but smaller mean $\sigma_{C/A}$ than the low- and intermediate-$C_{95}$ blue subsamples. In line with the Bayesian inference, the $q$ distribution of high-$C_{95}$ subsample is skewed to smaller values, and significantly different (K-S test $p$-values $<$ 0.05) from the low- and intermediate-$C_{95}$ subsamples. We note that there is no significant difference ($\textless 0.02$ dex) of median masses between the three $C_{95}$-defined subsamples.

  \par
  The sample size of red galaxies is too small to be further divided for a reliable 3D shape inference, so we divide the red galaxies into two halves based on $C_{95}$ and only show their $q$ distributions in Figure \ref{fig:red}. It is clear that the high-$C_{95}$ red galaxies have a $q$ distribution that is skewed to smaller $q$ than do the low-$C_{95}$ red galaxies, similar to the findings for the blue galaxies. The significance of the difference in the $q$ distributions is corroborated by the K-S test ($p$-value  = 0.006).

\setlength{\tabcolsep}{1.1em}
\renewcommand{\arraystretch}{1.1}
\begin{table*}[]
\centering
\caption{$\mu_{B/A}$, $\mu_{B/A}$ and their 16th-84th inter-percentile range for different subsamples analyzed in this work.}

\begin{tabular}{llccccl}
\hline
subsample & \multicolumn{2}{c}{$C_{95}$ \textgreater 2.62} & \multicolumn{2}{c}{2.26 \textless $C_{95}$ \textless 2.62} & \multicolumn{2}{c}{$C_{95}$ \textless 2.26} \\
\cmidrule(l{3pt}r{3pt}){2-3} \cmidrule(l{3pt}r{3pt}){4-5} \cmidrule(l{3pt}r{1pt}){6-7} 
& {$\mu_{B/A}$} & {$\mu_{C/A}$} & {$\mu_{B/A}$} & {$\mu_{C/A}$} & {$\mu_{B/A}$} & {$\mu_{C/A}$} \\
\cmidrule(l{3pt}r{3pt}){1-1} \cmidrule(l{3pt}r{3pt}){2-3} \cmidrule(l{3pt}r{3pt}){4-5} \cmidrule(l{3pt}r{1pt}){6-7}
$9.05 \textless \log(M_{\star}/M_{\odot}) \textless 9.80$$^{a}$ & $0.984_{-0.014}^{+0.011}$ & $0.434_{-0.017}^{+0.018}$ & $0.980_{-0.018}^{+0.013}$ & $0.473_{-0.016}^{+0.018}$ & $0.947_{-0.022}^{+0.033}$ & $0.502_{-0.017}^{+0.010}$ \\
$8.74 \textless \log(M_{\star}/M_{\odot}) \textless 9.05$$^{b}$ & $0.953_{-0.049}^{+0.045}$ & $0.406_{-0.014}^{+0.014}$ & $0.961_{-0.018}^{+0.027}$ & $0.486_{-0.017}^{+0.017}$ & $0.979_{-0.018}^{+0.015}$ & $0.466_{-0.016}^{+0.017}$ \\
$7.20 \textless \log(M_{\star}/M_{\odot}) \textless 8.74$$^{c}$ & $0.937_{-0.046}^{+0.063}$ & $0.418_{-0.017}^{+0.092}$ & $0.950_{-0.038}^{+0.050}$ & $0.457_{-0.017}^{+0.032}$ & $0.980_{-0.019}^{+0.013}$ & $0.462_{-0.017}^{+0.020}$ \\

blue$^{d}$ & $0.958_{-0.049}^{+0.050}$ & $0.411_{-0.007}^{+0.007}$ & $0.986_{-0.015}^{+0.010}$ & $0.462_{-0.011}^{+0.015}$ & $0.962_{-0.028}^{+0.026}$ & $0.475_{-0.034}^{+0.036}$ \\

blue \& central$^{e}$ & $0.979_{-0.011}^{+0.011}$ & $0.413_{-0.010}^{+0.011}$ & $0.981_{-0.017}^{+0.013}$ & $0.461_{-0.021}^{+0.016}$ & $0.973_{-0.019}^{+0.027}$ & $0.450_{-0.016}^{+0.015}$ \\
blue \& satellite$^{f}$ & $0.962_{-0.048}^{+0.036}$ & $0.415_{-0.021}^{+0.020}$ & $0.980_{-0.018}^{+0.018}$ & $0.462_{-0.015}^{+0.015}$ & $0.968_{-0.022}^{+0.022}$ & $0.463_{-0.015}^{+0.015}$ \\

central$^{g}$ & $0.933_{-0.043}^{+0.056}$ & $0.405_{-0.012}^{+0.013}$ & $0.978_{-0.017}^{+0.021}$ & $0.426_{-0.008}^{+0.007}$ & $0.975_{-0.038}^{+0.018}$ & $0.457_{-0.021}^{+0.015}$ \\
satellite$^{h}$ & $0.964_{-0.055}^{+0.035}$ & $0.433_{-0.027}^{+0.028}$ & $0.971_{-0.027}^{+0.021}$ & $0.453_{-0.017}^{+0.019}$ & $0.979_{-0.017}^{+0.024}$ & $0.485_{-0.016}^{+0.015}$ \\

\hline
\end{tabular}

      \begin{tablenotes}
      \small
      \item[a] (a-c): The result is also plotted in Figure \ref{fig:MASS_CI}; (d): The blue subsample is defined according to the NUV--r vs. mass diagram (Figure \ref{fig:scatter}), and the result is also plotted in Figure \ref{fig:blue_CI}; (e-f): Central and satellite galaxies that fall into the blue subsample (also plotted in Figure \ref{fig:blue_envirCI}); (g-h): Central and satellite subsamples, without regard for colors (also plotted in Figure \ref{fig:envir_CI})
      \end{tablenotes}

\label{table1}
\end{table*}

\setlength{\tabcolsep}{1.1em}
\renewcommand{\arraystretch}{1.1}
\begin{table*}[]
\centering
\caption{$\sigma_{B/A}$, $\sigma_{B/A}$ and their 16th-84th inter-percentile range for different subsamples analyzed in this work.}
\begin{tabular}{llccccl}
\hline
subsample & \multicolumn{2}{c}{$C_{95}$ \textgreater 2.62} & \multicolumn{2}{c}{2.26 \textless $C_{95}$ \textless 2.62} & \multicolumn{2}{c}{$C_{95}$ \textless 2.26} \\
\cmidrule(l{3pt}r{3pt}){2-3} \cmidrule(l{3pt}r{3pt}){4-5} \cmidrule(l{3pt}r{1pt}){6-7} 
& {$\sigma_{B/A}$} & {$\sigma_{C/A}$} & {$\sigma_{B/A}$} & {$\sigma_{C/A}$} & {$\sigma_{B/A}$} & {$\sigma_{C/A}$} \\
\cmidrule(l{3pt}r{3pt}){1-1} \cmidrule(l{3pt}r{3pt}){2-3} \cmidrule(l{3pt}r{3pt}){4-5} \cmidrule(l{3pt}r{1pt}){6-7}
$9.05 \textless \log(M_{\star}/M_{\odot}) \textless 9.80$$^{a}$ & $0.018_{-0.013}^{+0.019}$ & $0.181_{-0.019}^{+0.019}$ & $0.025_{-0.017}^{+0.024}$ & $0.191_{-0.018}^{+0.021}$ & $0.052_{-0.029}^{+0.030}$ & $0.173_{-0.015}^{+0.019}$ \\
$8.74 \textless \log(M_{\star}/M_{\odot}) \textless 9.05$$^{b}$ & $0.110_{-0.025}^{+0.027}$ & $0.110_{-0.014}^{+0.016}$ & $0.045_{-0.042}^{+0.026}$ & $0.173_{-0.016}^{+0.021}$ & $0.037_{-0.036}^{+0.036}$ & $0.169_{-0.016}^{+0.018}$ \\
$7.20 \textless \log(M_{\star}/M_{\odot}) \textless 8.74$$^{c}$ & $0.177_{-0.047}^{+0.065}$ & $0.108_{-0.012}^{+0.015}$ & $0.120_{-0.036}^{+0.035}$ & $0.153_{-0.018}^{+0.018}$ & $0.023_{-0.020}^{+0.032}$ & $0.166_{-0.015}^{+0.023}$ \\

blue$^{d}$ & $0.170_{-0.029}^{+0.028}$ & $0.110_{-0.006}^{+0.010}$ & $0.012_{-0.009}^{+0.020}$ & $0.165_{-0.009}^{+0.015}$ & $0.052_{-0.049}^{+0.032}$ & $0.166_{-0.025}^{+0.011}$ \\

blue \& central$^{e}$ & $0.010_{-0.007}^{+0.013}$ & $0.102_{-0.008}^{+0.009}$ & $0.025_{-0.022}^{+0.028}$ & $0.162_{-0.014}^{+0.016}$ & $0.037_{-0.033}^{+0.035}$ & $0.152_{-0.018}^{+0.019}$ \\
blue \& satellite$^{f}$ & $0.115_{-0.040}^{+0.060}$ & $0.135_{-0.015}^{+0.025}$ & $0.029_{-0.020}^{+0.038}$ & $0.155_{-0.013}^{+0.016}$ & $0.054_{-0.026}^{+0.026}$ & $0.165_{-0.015}^{+0.018}$ \\

central$^{g}$ & $0.121_{-0.043}^{+0.050}$ & $0.109_{-0.011}^{+0.013}$ & $0.035_{-0.018}^{+0.016}$ & $0.168_{-0.015}^{+0.012}$ & $0.058_{-0.051}^{+0.055}$ & $0.190_{-0.021}^{+0.022}$ \\
satellite$^{h}$ & $0.154_{-0.061}^{+0.039}$ & $0.130_{-0.015}^{+0.029}$ & $0.069_{-0.056}^{+0.037}$ & $0.160_{-0.016}^{+0.023}$ & $0.024_{-0.022}^{+0.022}$ & $0.176_{-0.015}^{+0.016}$ \\

\hline
\end{tabular}

\begin{tablenotes}
      \small
      \item[a] (a-c): The result is also plotted in Figure \ref{fig:MASS_CI}; (d): The blue subsample is defined according to the NUV--r vs. mass diagram (Figure \ref{fig:scatter}), and the result is also plotted in Figure \ref{fig:blue_CI}; (e-f): Central and satellite galaxies that fall into the blue subsample (also plotted in Figure \ref{fig:blue_envirCI}); (g-h): Central and satellite subsamples, without regard for colors (also plotted in Figure \ref{fig:envir_CI})
\end{tablenotes}

\label{table2}
\end{table*}

  \subsection{ Dependence on Local Environment}
  \par
  Here we explore the intrinsic shape of central galaxies and satellite galaxies separately. Results for the blue subsamples and the whole sample are plotted in Figures \ref{fig:blue_envirCI} and \ref{fig:envir_CI}, respectively. The results are also given in Tables \ref{table1} and \ref{table2}. Similar to the finding in preceding sections, high-$C_{95}$ subsamples have flatter intrinsic shapes (i.e., smaller mean $\mu_{C/A}$) than the low- and intermediate-$C_{95}$ subsamples, regardless of being central or satellite.\ In addition, for given $C_{95}$, there is no significant difference of intrinsic axis ratios and their standard deviations between the central and satellite subsamples. The above mentioned $C_{95}$-dependent differences of the intrinsic shape are also reflected in the $q$ distributions and the K-S test thereof.

  \subsection{Dependence on galactocentric radii where the axis ratios are measured}
  \par
  There is a concern on whether the above difference between high- and low-concentration subsamples reflects a real shape difference or merely an illusion resulted from the way that the ellipticity measurement is chosen. Particularly, if the high-concentration galaxies have systematically different radial variations of stellar mass-to-light ratio compared to the lower-concentration galaxies, our choice of ellipticity measured near a fixed surface brightness level ($\lesssim$ 27 mag/arcsec$^{2}$) would result in a difference in surface mass density levels where the intrinsic shape is inferred. To verify whether this potential bias could affect the main results presented in this section, we infer the intrinsic mean axis ratios of each subsample based on average ellipticities measured near $\leq$ 26 mag/arcsec$^{2}$ (i.e. smaller galactocentric radii than the default choice), and then compare the intrinsic mean axis ratios of high-concentration subsamples inferred from the ellipticities measured near 26 mag/arcsec$^{2}$ to that of lower-concentration subsamples inferred from ellipticities measured near 27 mag/arcsec$^{2}$, and vice versa. We find that (not shown here) the high-concentration subsamples have slightly lower intrinsic mean $\mu_{C/A}$ than do low-concentration subsamples in virtually all cases. This experiment corroborates the conclusion that the high-concentration subsamples have slightly lower or at most similar intrinsic thickness than do the lower-concentration subsamples.
  
 \section{discussion}\label{sec:discussion}
   \par
     It has been established by early numerical simulations that mergers of {relatively massive} disk galaxies result in remnants with much denser centers (due to torque-driven gas inflow and subsequent central starburst; e.g., \citealt{1991ApJ...370L..65B}) but more extended outskirts (due to violent relaxation; e.g., \citealt{1981MNRAS.197..179G}) than their progenitors (e.g., \citealt{1996ApJ...471..115B}). The structural properties of remnants resulted from major mergers resemble nearby classical elliptical galaxies. Similar results were reported {by some simulations} of mergers of dwarf galaxies (e.g., \citealt{2008MNRAS.388L..10B}). In particular, the radial mass density profiles of merger remnants, as compared to the progenitor galaxies, are characterized by relatively high S\'{e}rsic indices, which is correlated with the concentration parameters as used in this work. The change of the radial profiles is accompanied with a transformation of the intrinsic shapes (e.g., \citealt{2000MNRAS.312..859S}). 
     
     {Simulations for relatively massive disk galaxies suggest that}, while the merger remnants tend to have larger triaxiality than their progenitors in a statistical sense, the exact shape depends on the collision geometry and the progenitor properties, (e.g.\ mass ratios, morphology and gas fraction). In particular, {gas-rich (dissipational) mergers result in remnants that are closer to oblate, whereas gas-poor (dissipationless) mergers leave remnants that have a larger triaxiality and are closer to prolate (e.g., \citealt{2006ApJ...650..791C, 2021A&A...647A..95P}). Moreover, a merger remnant may eventually evolve into a new disk galaxy if sufficient gas remains long after the merger event \citep[e.g.][]{Athanassoula16}.}

     \par
     {Late-type dwarf galaxies generally have a much higher gas fraction than their high-mass counterparts, especially in low density environment. Therefore, we may expect gas-rich dwarf-dwarf merger remnants to be as oblate as ordinary late-type dwarfs. {Nevertheless, some recent simulations \citep[e.g.][]{Dicintio2019} suggest that the connection between merger orbital configuration and the properties of the remnant becomes weaker for lower mass galaxies}. This makes it non-trivial to distinguish between merger remnants and ordinary dwarfs.} Indeed, we find that our sample of galaxies is close to be oblate ($\mu_{B/A}$ $\sim$ 0.9--1) on average, regardless of concentration indices. On the other hand, our finding that the highest-concentration dwarfs tend to be slightly ($\Delta(\mu_{C/A})$ $\sim$ 0.05--0.1) but significantly thinner, instead of thicker, than lower-concentration dwarfs seems to be inconsistent with {the classical} merger scenario for high-concentration {galaxies}. {Recent observational studies of interacting dwarf pairs \citep[e.g.][]{2015ApJ...805....2S, 2017ApJ...846...74P,2020AJ....159..103K} and dwarf merger remnants \citep{Zhang2020a,Zhang2020b} imply that dwarf-dwarf mergers may not simply be scale-down versions of their massive counterparts. Given the limited knowledge of the general outcome of dwarf-dwarf mergers at this time, we cannot draw a solid conclusion about the relevance of dwarf-dwarf mergers in the formation of high-concentration dwarfs.}
  
  \par
  Besides mergers, spin of a dark matter halo is another important factor that may determine the morphology of galaxy formed near the halo center. Particularly, \cite{2017MNRAS.467.3083R} studied the connection of assembly histories and halo spin with galaxy morphologies based on the Illustris simulations, and found that morphology of massive galaxies with $M_{\star}$ $\geq $ $10^{11}$ $M_{\odot}$ is primarily correlated with merger histories, whereas for galaxies with $M_{\star}$ $\textless$ $10^{10}$ $M_{\odot}$, the morphology is primarily correlated with halo spin (see however \cite{2022MNRAS.tmp..808R} based on the IllustrisTNG simulations). Lastly, violent disk instabilities have also been put forward as a viable mechanism to form high-concentration dwarfs such as blue compact dwarfs \citep{2012ApJ...747..105E}.\ Such violent disk instabilities may be triggered by gas inflow from the cosmic web.

  \section{Conclusions} \label{sec:summary}
    \par
    In this work, we assemble a large sample of nearby low-mass galaxies that are redshift-confirmed by SDSS and GAMA surveys and have relatively deep broadband optical images from the HSC-SSP survey. The sample covers a redshift range from 0.005 to 0.2 and a stellar mass range of $10^{7.5}$ to $10^{10.0}$ $M_{\odot}$. We measure the apparent axis ratios of these galaxies at surface brightness levels of $\sim$ 27.0 mag/$\rm arcsec^{2}$ that are beyond the central active star-forming regions. In a Bayesian framework, we use the apparent axis ratios to infer the average intrinsic shapes (parameterized by the intrinsic axis ratios B/A and C/A of triaxial ellipsoids, where A, B and C represents the major, intermediate and minor axis respectively) of our galaxies. We carry out a comparative study of the intrinsic shape of subsamples defined by radial light concentration ($C_{95}$ = $R_{90}/R_{50}$), stellar masses, star formation status (based on NUV--$r$ colors), and local environments (being central or satellite galaxies). {The main goal of this work is to test if there is any correlation between concentration index and intrinsic shape of dwarfs, which may shed light on the formation mechanism of high-concentration dwarf galaxies}.
    
    \par
    We find that different subsamples of dwarf galaxies have mean intrinsic shapes that are similarly close to be oblate (mean $\mu_{B/A}$ $\sim$ 0.9--1), regardless of $C_{95}$, stellar masses, star formation status and local environments. However, dwarfs with the highest $C_{95}$ tend to be slightly flatter (i.e., smaller mean $\mu_{C/A}$) than those with relatively low $C_{95}$, regardless of the stellar mass, star formation activity and local environment. This means that the highest-$C_{95}$ galaxies tend to have lower triaxiality (or thinner) than low-$C_{95}$ galaxies, which appears to be inconsistent the expectation of classic merger scenario. {But given the lack of a complete understanding of dwarf-dwarf mergers, we cannot draw a definite conclusion about the relevance of merger in the formation of high-$C_{95}$ dwarfs.}\ Other factors, such as halo spin and violent disk instabilities, may also play important roles in the formation of high-concentration dwarf galaxies.

\begin{acknowledgements}
  \par 
  We thank the anonymous referee for the very helpful comments that lead to a significant improvement of the manuscript. We acknowledge support from the China Manned Space Project (No. CMS-CSST-2021-A07, CMS-CSST-2021-B02),  the NSFC grant (No. 11421303, 11973038, 11973039, 12122303, 12233008), and the CAS Pioneer Hundred Talents Program, the Strategic Priority Research Program of Chinese Academy of Sciences (Grant No. XDB 41000000) and the Cyrus Chun Ying Tang Foundations. Z.S.L. acknowledges the support from China Postdoctoral Science Foundation (2021M700137).
\end{acknowledgements}

\bibliography{apollo_apj}{}
\bibliographystyle{aasjournal}

\end{document}